# Influence of Superconductor Dirtiness on the SNSPD Sensitivity-Bandwidth Trade-off

Souvik Haldar, Yash Sharma, Krishna B. Balasubramanian[#]

*Department of Material Science and Engineering, Indian Institute of Technology Delhi, New Delhi 110016, India*

[#]Email (Corresponding author): bkrishna@mse.iitd.ac.in

## Abstract

Practical superconducting nanowire single photon detectors (SNSPDs) demonstrate a strong trade-off between detection sensitivity and the reset time. Often, there are wide variations in sensitivity and response times from SNSPDs of the same superconducting material. Here, using detailed physical models, we show that the dirtiness in a superconductor enforces a sensitivity and bandwidth trade-off in all practical devices. More importantly, a certain degree of dirtiness is a necessary requirement for achieving single photon detection. Under typical bias conditions close to the transition setpoints, the minimum number of photons required to register a voltage pulse decreases by the dirtiness parameter (Ioffe-Regel parameter) and the reset time of SNSPD increases by the same dirtiness parameter, thereby giving a constant value for the sensitivity-bandwidth product. The constant is weakly modified by biasing current and the temperature. Since dirtiness in the superconducting nanowire is a physically controllable parameter with an important bearing on the final response of an SNSPD, this work opens new opportunities to develop SNSPD devices with engineered sensitivity-bandwidth setpoint as dictated by an application.

## Main

The efficiency and timing response of a superconducting nanowire single photon detector (SNSPD) depends on the dynamics of hotspot formation and dissipation upon a photon incidence. An ideal SNSPD should respond to a unit photon incidence with vanishingly small reset times thereby allowing for both high sensitivity and high detection speeds. Hence, there is an increasing interest to improve photon sensitivity of a practical SNSPD while also promoting faster hotspot dynamics [1–4]. However, these parameters are closely related in a SNSPD with a persistent trade-off. The reset time of a SNSPD is a parameter derived from the time taken for the hotspot to dissipate the heat from photon incidence into the substrate and environment [5]. The thermal boundary resistance ($\tau_{esc}$), which determines the rate of heat transfer into the substrate, depends on the choice of superconductor and the substrate [6]. The heat diffusion (electronic ($D_e$) and phonon ($D_{ph}$)) coefficients and the diffusion time scales $t_D \approx \frac{w^2}{4D_e}$ influence the rate of heat transfer. Smaller reset times require larger diffusion coefficients. On similar lines, the photon sensitivity is the minimum number of photons required to completely break the superconductivity over the entire width of the wire [7]. For this to happen, the thermalization of the photoexcited carriers by subsequent Cooper-pair excitations (determined by electron-phonon relaxation times, $\tau_{e-ph}$) should be faster than the thermal diffusion time scales [8]. This makes superconductors with smaller electron heat diffusion coefficient, $D_e$, to be better choices for SNSPD, as the smaller $D_e$ (higher diffusion times, $t_D \approx w^2/4D_e$ where $w$ is the width of the wire) reduces energy dissipation, consequently, improving detection sensitivity [8]. All the parameters mentioned above are strong functions of not only the choice of the superconducting material but also its 'dirtiness' (that includes effects from crystalline defects, dopants, surface effects and more) [9–11]. Thus, the route to improve/control the photon sensitivity and detection rate of SNSPs lies in the ability to estimate the crucial role played by the choice of the material, its defect constitution and its net effect on the physical properties the influence the SNSPD response. The aspects are further complicated by the nanoscale patterning steps that help realize the wire, since they introduce defects and material property changes that influences the device response appreciably [1].

The relation between two important length scales in a superconductor (mean free path, $l$, and BCS coherence length, $\xi_0$) gives two different classes: (i) clean superconductor ($l > \xi_0$) and (ii) dirty superconductor when otherwise [12]. Conventionally, dirtiness is quantified by using the Ioffe-Regel parameter ($k_F l$) where $k_F$ is the Fermi wavevector. (i) $k_F l > 1$ implies the metallic clean limit, (ii) $k_F l < 1$ implies the insulating region and (iii) $k_F l \sim 1$ implies Mott limit [13,14]. Here for simplicity, we refer it by $D = 1/k_f l$. While some extent of dirtiness is inherently present in all bulk superconductors, dirtiness is also introduced by structural and chemicals methods [12]. Mechanical stresses during the fabrication process of an SNSPD introduce crystal imperfections by means of vacancies, dislocations, and grain boundaries [15]. External additions introducing electron doping, localized magnetic moment, surface contamination also makes the SNSPD dirty [16]. The order of dirtiness, $k_F l$, affects different superconducting properties such as critical temperature ($T_c$), normal state resistivity ($\rho_N$), $D_e$, critical current density ($J_{c0}$), and heat capacity ($C_e$), which crucially influence the evolution of hotspot in a SNSPD [17]. Several experimental reports are available that demonstrate the influence of dirtiness. Different amorphous superconductors with different dirtiness levels such as WSi, MoSe, MoSi, and NbSi and disordered superconductor NbReN also showed lower jitter, faster, and near unit detection efficient SNSPD [18–20]. In 36 nm thick NbReN-SNSPD, where $k_F l$ is found to be 1.3, quasiparticle relaxation time has been found to be around 300 ps [20]. Therefore, the effect of dirtiness on the photon resolution and reset time of SNSPD is a common experimental fact. However, to the best of our knowledge, no definitive numerical treatment is present that correlates the effect of dirtiness and the performance of the SNSPD. As the SNSPD is rapidly replacing conventional photon detectors in several applications due to its extreme efficiency and low dark count rates, a reliable model that can correlate the material properties (particularly dirtiness in a superconductor) and the eventual photo response is a crucial requirement.

In this manuscript, we have developed a theoretical model based on modified two-temperature formalism by incorporating the effect of joule heating, photon pulse shapes, and the influence of dirtiness on the superconducting thermodynamic variables on the photon resolution and the hotspot dynamics of an SNSPD. We calculate the complete photo-response of an SNSPD (the final magnitude and the shape of the output voltage) with different order of dirtiness and estimate the minimum photon count required to register a voltage pulse. As a representative example, we study the impact of dirtiness on an NbN SNSPD. NbN is an *s*-wave superconductor with higher transition temperature ~ 16 K, short coherence length < 5 nm, and large penetration depth ~ 200 nm, which allows fabrication of thinner nanowires [21,22]. However, even the 'pristine' NbN with close to ideal superconducting transition is dirty ($l < \xi_0$). We vary the nanowire $D$ from metallic limit ($D = 0.15$) to insulating limit ($D = 1.0$) and irradiate it under a Gaussian photon pulse ($\lambda = 1550$ nm) with controllable photon count. We observe that in a 'dirty' superconductor with a Ioffe-Regel parameter $1/D$, the minimum number of photons required to register a pulse is reduced by $D$ times $\left(n_p^*(D) = \frac{n_{p,clean}^*}{D}\right)$, and the reset time increases approximately by $D$ folds ($\tau_{reset}(D) = \tau_{reset}^{clean} * D$). Thus, giving a general tradeoff $n_p^* \times \tau_{reset} \cong C$, wherein $C$ is found to be a weak function of the bias current. The high photon sensitivity and shorter reset time of SNSPD have their different applications. While a device with a high photon sensitive device typically has longer reset time. The longer reset time results in high jitter which eventually hinders fast counting rate. On the contrary, shorter reset time promotes faster counting rate which lacks higher photon sensitivity. Our calculation predicts that with engineered dirtiness, the NbN SNSPD can be used for applications ranging from ultrasensitive (but slow) to ultrafast by less sensitive photon detection at appropriate biasing conditions. We believe these results can let various device engineers suitably modify an SNSPD for an optimized sensitivity-detection rate setpoint.

The photo-response of an SNSPD under gaussian photon pulse, having length ($L$), width ($w$) and thickness ($t$) is investigated for different levels of dirtiness using the following modified two-temperature models. The exact details of the equations are described elsewhere.

$$C_e(T_e)\frac{\partial T_e}{\partial t} - k_e(T_e)\nabla^2 T_e + \frac{C_e(T_e)}{\tau_{e-ph}(T_e)}(T_e - T_{ph}) = f_p(x,t) + f_J(t) \dots (1)$$

$$C_{ph}(T_{ph})\frac{\partial T_{ph}}{\partial t} - k_{ph}(T_{ph})\nabla^2 T_{ph} - \frac{C_e(T_e)}{\tau_{e-ph}(T_e)}(T_e - T_{ph}) + \frac{C_{ph}(T_{ph})}{\tau_{esc}}(T_{ph} - T_{sub}) = 0 \dots (2)$$

Where, $T_e, T_{ph}$ and $T_{sub}$ are the electron, phonon and substrate temperatures respectively. $C_e, C_{ph}, k_e = D_e C_e$ and $k_{ph} = D_{ph} C_{ph}$ are the electron heat capacity, phonon heat capacity, electron thermal conductivity and phonon thermal conductivity respectively. $D_e$ and $D_{ph}$ are the electron and phonon diffusion coefficients respectively. The heat exchange between the electron and phonon sub-systems is expressed via the electron-phonon coupling time $\tau_{e-ph}$. The heat loss to the substrate is inversely proportional to the escape time of the non-equilibrium phonon to the substrate, $\tau_{esc}$. $f_p(x,t)$ is the Gaussian photon pulse and $f_J(t)$ is the Joule heating. The dirtiness affects the superconducting parameters, which govern the hotspot dynamics. These effects of dirtiness, $1/D = k_F l$, on the superconducting parameter of NbN are given in the *Supplementary material Section 1* [17,23–33].

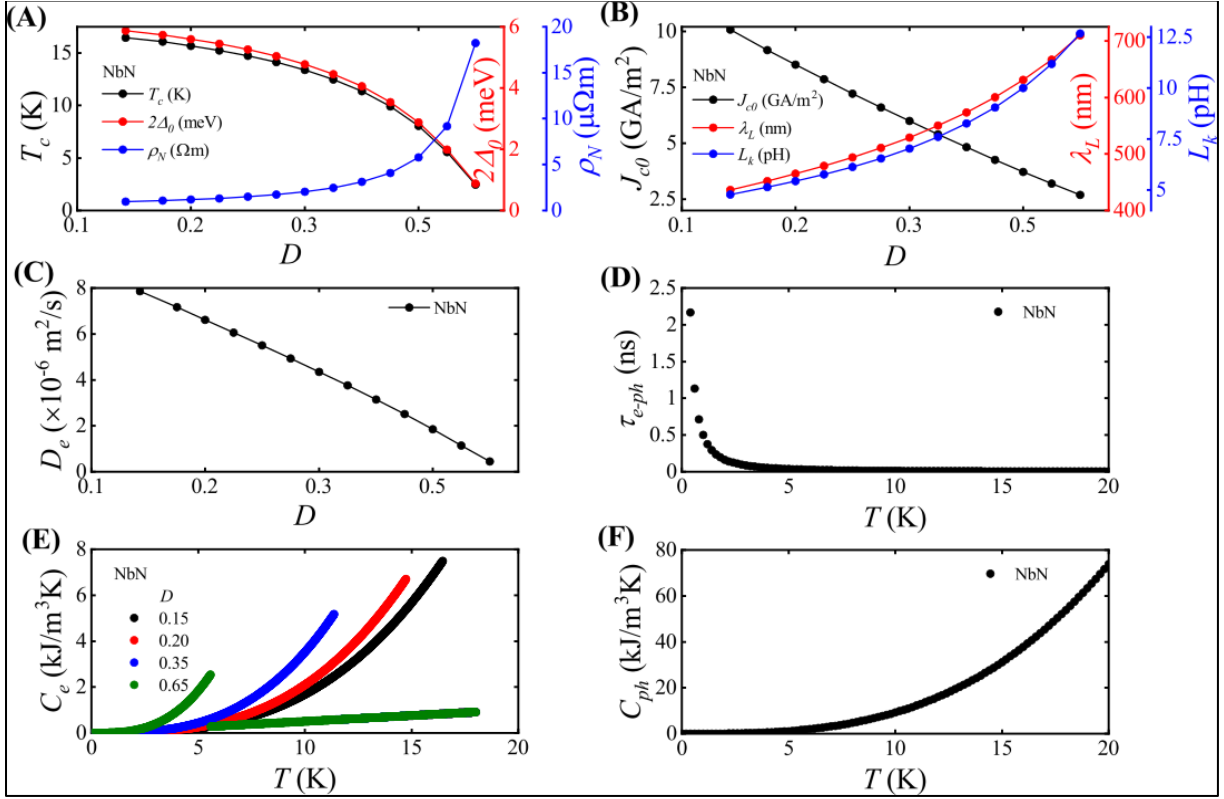

**Figure 01:** **(A)** Variation of transition temperature, $T_c$, superconducting order parameter, $2\Delta_0$, and normal state resistivity, $\rho_N$, with different order of dirtiness, $D$. **(B)** Variation of zero temperature critical current density, $J_{c0}$, London Penetration depth, $\lambda_L$, and kinetic inductance of the nanowire, $L_k$, with different order of dirtiness, $D$. **(C)** Variation of electron diffusion coefficient, $D_e$, with different order of dirtiness, $D$. **(D)** Variation of electron-phonon coupling time, $\tau_{e-ph}$, as a function of temperature, $T$. **(E)** Variation of electron heat capacity, $C_e$, as a function of temperature, $T$, for different level of dirtiness, $D$. **(F)** Variation of phonon heat capacity, $C_{ph}$, as a function of temperature, $T$.

Dirtiness influences transition temperature, which in-turn affects the order parameter, whose relation for NbN is reported to be $2\Delta(T) = 4.22 k_B T_c \tanh \tanh \left(1.74 \sqrt{\frac{T_c}{T} - 1}\right)$. The variation of order parameter at zero temperature ($2\Delta_0$) along with transition temperature, $T_c$, and normal state resistivity, $\rho_N$, are plotted in **Fig. 01(A)** for different level of dirtiness, $k_F l$. In the 'cleaner' NbN, we observe the experimentally measured maximal transition temperature, $T_{c,max} = 16.46$ K and $2\Delta_0 = 5.87$ meV. As the superconductor becomes dirty, $T_c$ and $2\Delta_0$ decreases monotonically and at the Mott limit, $1/D = k_F l = 1$, $T_c$ and $2\Delta_0$ decreases drastically from the clean limit by seven and three folds respectively. Experimentally transition width, $\Delta T_c$ increases by the same order as the $D$ value ($\Delta T(D) \cong 1.2 * D * \Delta T_{clean}$). From experiments, resistivity shows non-metallic character with increasing dirtiness as seen in **Fig. 01(A)**. When the $D$ value approaches to Mott limit ($1/D = k_F l = 1$) NbN shows a superconductor to insulator transition [34]. In **Fig. 01(B)**, we show the variation of zero temperature critical current density, $J_{c0}$, London penetration depth, $\lambda_L$, and kinetic inductance of the nanowire, $L_k$. For the $D$ value variations we have taken, $J_{c0}$ falls off by four folds and $L_k$ increases by 2.5 times. London penetration depth, $\lambda_L$, also increases 1.75 times with the extent of dirtiness increase taken here. $L_k$ is proportional to $\lambda_L$ (Please refer to *TABLE 1.1 of Supplementary material Section 1*). We have shown the variation of electron diffusion coefficient, $D_e$, with dirtiness in **Fig. 01(C)**. The minimum $D_e$ is found to be $0.44 \times 10^{-6}$ m²/s for the dirtier superconductor which is more than an order of magnitude smaller than $D_e$ observed in the cleaner one. Electron-phonon coupling time, $\tau_{e-ph}$ is found to be proportional to $T^{-n}$. From the experimental data, the exponent, $n$, is found to be 1.6 [32]. **Fig. 01(D)** represents the variation of $\tau_{e-ph}(T)$. For an ultrathin NbN film (2.5 nm thin) $\tau_{e-ph}(T)$ does not get affected by disorder; which is also observed experimentally [35]. The dirtiness affects the variation of electron heat capacity in superconducting state, $C_{es}$; though above the transition temperature in the normal state heat capacity, $C_{en}(T)$ shows regular linear variation as seen in **Fig. 01(E)**. A typical discontinuity in the specific heat at $T = T_c$ is observed for all the $D$ values considered here. The variation of phonon heat capacity as a function of temperature is shown in **Fig. 01(F)**. Standard $T^3$ dependence has been observed in the variation of $C_{ph}(T)$ below the Debye temperature, $\Theta_D$. $C_{ph}$ depends on the number density of lattice points in the system, with dirtiness the number density of the lattice points remains unaltered resulting no change in $C_{ph}(T)$.

We calculated the thermal distribution due to the photon incidence on the nanowire by solving the equations **1** and **2** simultaneously and plot the resultant temperature profile, $T_e(x,t)$, as a function of position, $x$, and time, $t$. For the sake of calculations, a Gaussian pulse of $n_p$ number of photons having 1550 nm wavelength is made to incident on the wire exchanging energy with the pairs. In the single photon limit, the photon transfers energy to a single calculation element. For the other cases, the photon pulse is made to take a Gaussian profile in both space and time. The photon pulse delivers its maximum power at $x = 50$ μm. The calculated temperature of the nanowire normalized by the switching temperature for a representative bias current of $I_b = 0.6 I_c$ and set substrate temperature of $T_{sub} \sim 0.9 T_{sw}$ is presented in the **Fig. 02(A)** and 0**2(B)** for two dirtiness levels, $D = 0.15$ and $0.5$ respectively. The wire is irradiated with a pulse of 30 photons, $(n_p = 30)$ for both the $D$ values and it exchanges the energy with a single calculation element of the nanowire. When the normalized temperature, $T_e/T_{sw}$, reaches 1, the superconductivity is lost and the material turns to normal state.

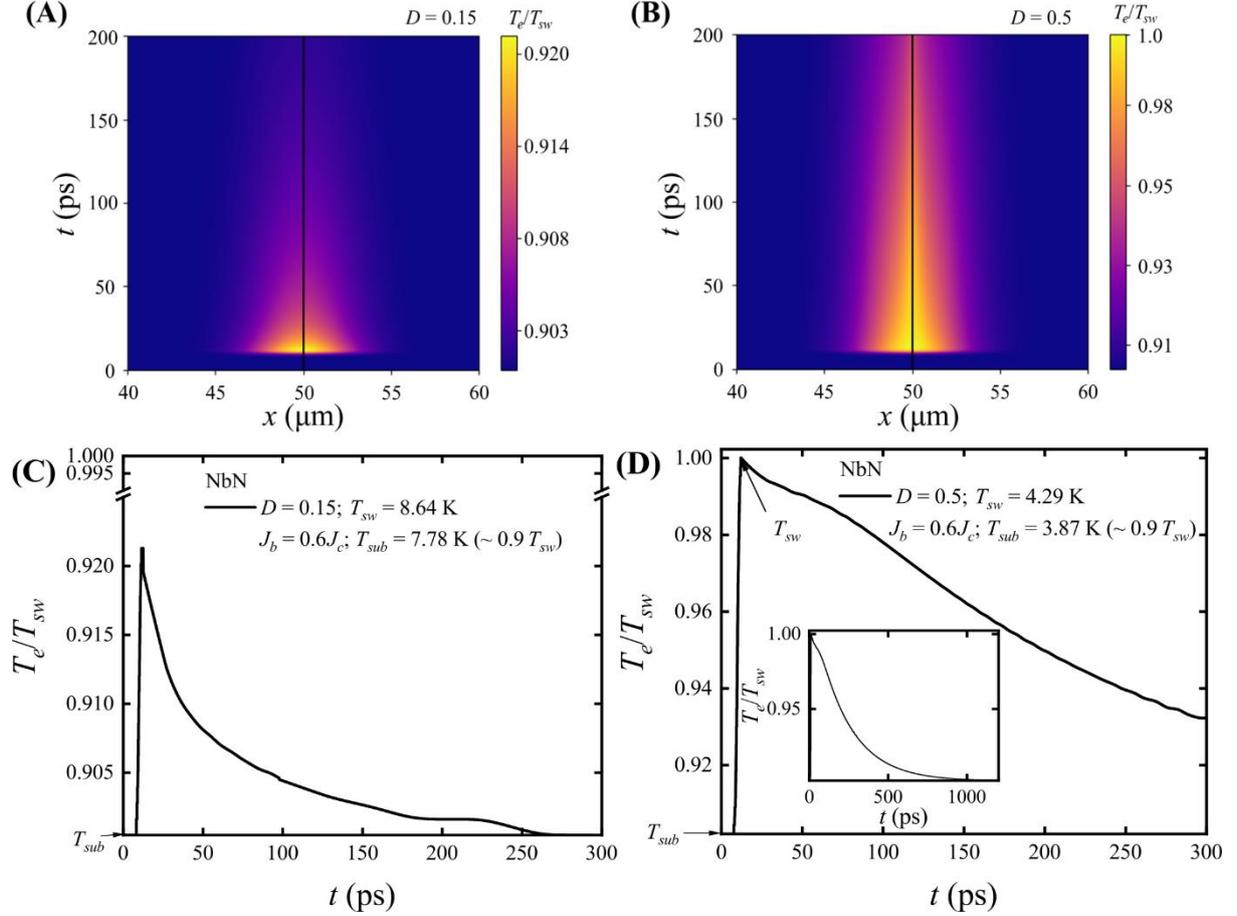

**Figure 02:** Variation of normalized temperature $T_e(x,t)/T_{sw}$ for different order of dirtiness, $D$ = **(A)** 0.15, and **(B)** 0.5, where $T_{sw}$ is the switching temperature respectively; Variation of normalized temperature, $T_e/T_{sw}$ as a function time, $t$, at $x = 50$ μm for $D$ = **(C)** 0.15, and **(D)** 0.5. The device was biased with $J_b/J_c = 0.6$ and at $T_{sub}/T_{sw} \sim 0.9$ for both the $D$ values.

There are distinct differences in the time-temporal response of the wires with different $D$ values. Quite clearly, the temperature never reaches to the switching temperature in the cleaner $D = 0.15$ case, while the superconductivity is broken for the same photon pulse in the dirtier $D = 0.5$ case. The time profile of the normalized temperature at the center of the hotspot is presented in **Fig. 02(C)** and **Fig. 02(D)**. While both the wires have similar rise times, the $D = 0.5$ wire had twice the temperature change of the cleaner $D = 0.15$ wire. However, a more striking difference is in the reset time. The reset time of the $D = 0.5$ wire is almost four times that of the cleaner one as can be noted from the inset of **Fig. 02(D)**. The potential from the hotspot formation is calculated from the equivalent circuit model as described elsewhere [36]. The details are presented in the *Supplementary material Section 2*.

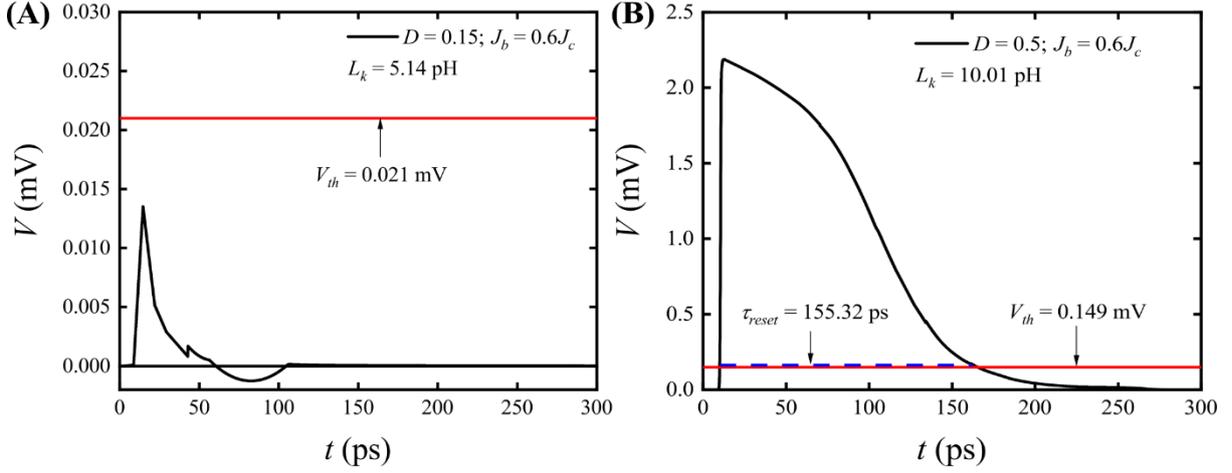

**Figure 03:** Variation of output voltage pulse, $V$, as a function of time, $t$, for $D =$ **(A)** 0.15 and **(B)** 0.5. The red horizontal line represents the thermal noise floor, $V_{th} = \sqrt{4k_B \Delta f R T}$ at the particular biasing condition.

The photon pulse creates a normal region in the SNSPD which in turns creates a detectable voltage pulse, $V(t)$. Variations of $V(t)$ for $D = 0.15$ and $0.5$ have been shown in **Fig. 03(A)** and **03(B)** respectively. The horizontal red lines in both the micrographs indicate the thermal noise, $V_{th} = \sqrt{4k_B T R \Delta f}$; which have values 0.021 mV and 0.149 mV for $D = 0.15$ and 0.5 respectively. The maximum output voltages, $V_{max}$, have been observed 0.014 mV and 2.19 mV for $D = 0.15$ and 0.5 respectively. Therefore, while the $D = 0.5$ SNSPD can resolve $n_p = 30$ pulse, under same biasing conditions, $D = 0.15$ does not register a measurable potential. Further calculations reveal that the $D = 0.15$ wire will register a measurable pulse with a pulse whose photon count is 2.5 times higher ($n_p = 75$).

We now calculate the minimum number of photons, $n_p^*$, for any given biasing conditions for different dirtiness values $D$. $n_p^*$ is obtained by calculating the potential profile and checking if the voltage peak appears above the minimum thermal noise limit. In **Fig. 04(A)**, we have shown the variation of $n_p^*$ as a function of $J_b/J_c$ for different $k_F l$. $n_p^*$ decreases monotonically (the wire becomes more sensitive) as the current density increases for all the $k_F l$. This is expected as the bias points get closer to the transition positions. At a constant $J_b$, the $n_p^*$ decreases as $D$ increases. Therefore, as dirtiness increases, the nanowire creates a hotspot and registers a potential for a pulse of smaller photon count, and hence, is more sensitive. The minimum number of photons, $n_p^*$, the NbN SNSPD can resolve falls drastically and approaches the single photon limit as the dirtiness increases to about $D = 0.5$ at $J_b/J_c = 0.9$. The ultimate single photon number resolution can be observed in NbN SNSPD at $J_b/J_c = 0.95$ with dirtiness $D = 0.5$.

The reset time, $\tau_{reset}$, after which the SNSPD is ready again for another photon detection can also be inferred from the potential plot. For the $D = 0.5$, $\tau_{reset}$ is calculated to be 155.32 ps. It has been experimentally observed that, for 500 μm long 4 nm thin NbN SNSPD was exposed under 950 nm wavelength pulse with 1 ps pulse width; the reset time was found to be about 350 ps [37]. 6.5 nm thick 90 nm wide NbN SNSPD under 1550 nm photon source was reported to have a reset time of about 358 ps [38]. The close proximity to many experimentally observed detector timings validates the model accuracy. We calculated the variation of $\tau_{reset}$ of the SNSPD as a function of biasing current, $J_b/J_c$, for different $D$ in **Fig. 04(B)**. $\tau_{reset}$ increases monotonically as $J_b$ increases for all the $D$. The minimum reset time, $\tau_{reset,min}$, has been observed to be 10.46 ps at $J_b/J_c = 0.05$ for $D = 0.15$; which is the least sensitive condition for the NbN SNSPD. Whereas, the maximum $\tau_{reset,max}$ has been observed to be almost two orders of magnitude higher than $\tau_{reset,min}$ at $J_b/J_c = 0.95$ for 3 times dirtier SNSPD ($D = 0.5$), which is the most sensitive condition for NbN SNSPD. It is important to mention that, at a constant biasing current $\tau_{reset}$ increases atleast 3 times with similar change in the dirtiness in the higher biasing current regime ($J_b/J_c \geq 0.8$).

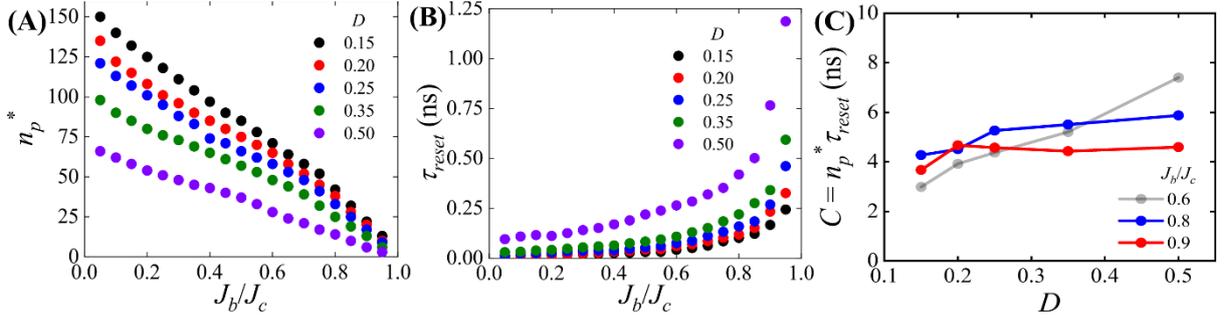

**Figure 04:** Variation of **(A)** reset time, $\tau_{reset}$, and **(B)** number of photons, $n_p^*$, needed to create an observable hotspot as a function of biasing current densities, $J_b/J_c$, for different $D$ values. The substrate temperature, $T_{sub}$, was always kept at $0.9T_{sw}$. **(C)** Variation of trade-off parameter, $C = n_p^* \times \tau_{reset}$, as a function of $D$ for different $J_b/J_c$.

It is quite clear that under commonly used bias conditions, ($J_b/J_c \geq 0.8$) for maximal sensitivity, the $n_p^*$ and $\tau_{reset}$ have contrasting influence of the $D$ values. This gives a trade-off between these two critical parameters. The product of the two $C = n_p^* \times \tau_{reset}$ as a function of $D$ for different $J_b/J_c$ values is presented in **Fig. 04(C)**. The approximately constant $C(D)$ at different biasing condition $J_b$ is also presented. As can be noted, at biasing conditions suitable for higher sensitivity ($J_b/J_c \geq 0.8$), near flat lines of the product enforces a strong trade-off as mentioned previously. The constant is a weak function of the bias current value. At lower biasing conditions ($J_b/J_c \leq 0.6$), the dirtiness has a stronger effect on critical photon count in comparison with the $\tau_{reset}$ showing a weaker correlation between the $n_p^*$ and $\tau_{reset}$ under such conditions.

We now discuss the reasoning behind the results observed. The reset time of SNSPD is proportional to (a) the rate of heat dissipation to substrate and (b) the kinetic inductance. The heat dissipation rate into the substrate is largely invariant for a chosen substrate. The kinetic inductance, $L_k$ (please see **Fig. 01(B)**) increases with dirtiness as $\lambda_L$ also increases. The reset time of NbN SNSPD increases as length of the SNSPD, $l$, increases. The reset time, $\tau_{reset}$, is roughly proportional to the ratio between $L_k$, $R(t)$ and $R_L$; $\tau_{reset} \propto \frac{L_k}{(R_L+R(t))}$. Therefore, with the increase of dirtiness, the reset time also increases. With the increase of dirtiness, electron thermal diffusion coefficient $D_e$ decreases; which hinders the diffusion of hotspot heat carried by the electrons. It favours concentration of the photon energy close to the incident spot thereby breaking more Cooper pairs and increasing local temperature further. Therefore, a smaller number of photons is needed to break the superconductivity, increasing the photon resolution of the 'dirtier' SNSPD. $\rho_N$ increases and $J_{c0}$ decreases as the dirtiness increases, which in turn decreases the Joule heating. The change of $n_p^*$ at higher biasing current $\left(\frac{J_b}{J_c} \geq 0.8\right)$ does not vary significantly with $D$. Smaller $D_e$ plays dominant role to determine the resolution of the SNSPD when $J_b < 0.8J_c$; above $J_b > 0.8J_c$ $D_e$ does not play dominant role; the reduced Joule heating rate hinders the drastic change of the $n_p^*$.

The generation of normal state in the SNSPD depends on the number of excess energised electrons ($N_e$) and the number of superconducting electrons in the equilibrium, $N_{se}$, in the hotspot. The $N_e$ and $N_{se}$ should satisfy the relation, $N_e/N_{se} \geq 1$ for detecting a successful photon pulse. $N_{se}(T) = n_{se}(T)\Delta x w d$, where $n_{se}(T)$ and $\Delta x$ are the temperature dependent conducting electron density and spatial variation of the hotspot respectively. $n_{se}$ can be calculated from the definition of London's penetration depth, $\lambda_L(T) = \sqrt{m_e/\mu_0 n_{se}(T)e^2}$, where $m_e$ and $e$ are the mass and charge of single electron respectively [40]. We assumed all the conducting electrons form Cooper pairs below the transition temperature, $T_c$. The variation of $\lambda_L$ as a function of temperature, $T$, follows $\lambda_L(T) = \lambda_L(T=0)/\sqrt{1-(T/T_c)^4}$. Therefore, the number of conducting electrons, $n_{se}(T)$, at a temperature, $T$, becomes $n_{se}(T) = n_{se}(T=0)(1-(T/T_c)^4)$. The superconducting electron density then become $N_{se}(T) = n_{se}(T=0)\Delta x w d(1-(T/T_c)^4)$. We can calculate the number of energised excess electrons $N_e$ from the following relation;

$$N_e(t) = \frac{\zeta f_p(t) + f_J(t)}{\Delta} \frac{\tau_r}{\tau_r - \tau_{e-ph}} \left[ \exp\left(-\frac{t}{\tau_r}\right) - \exp\left(-\frac{t}{\tau_{e-ph}}\right) \right] \tanh\left(\frac{\Delta x}{\sqrt{4\pi D_e t}}\right) \quad \ldots (3)$$

$\zeta, f_p(t), f_J(t)$ and $\Delta$ are the conversion efficiency, energy of the photon pulse having wavelength 1550 nm, Joule heating energy as a consequence of the onset of normal region and superconducting energy gap respectively. $\tau_r$ and $\tau_{e-ph}$ are the Cooper pair recombination time and electron-phonon coupling time respectively [41]. Assuming maximal conversion efficiency $\zeta = 1$, for NbN SNSPD, $\tau_r(\gg \tau_{e-ph})$ is found to be 1000 ps.

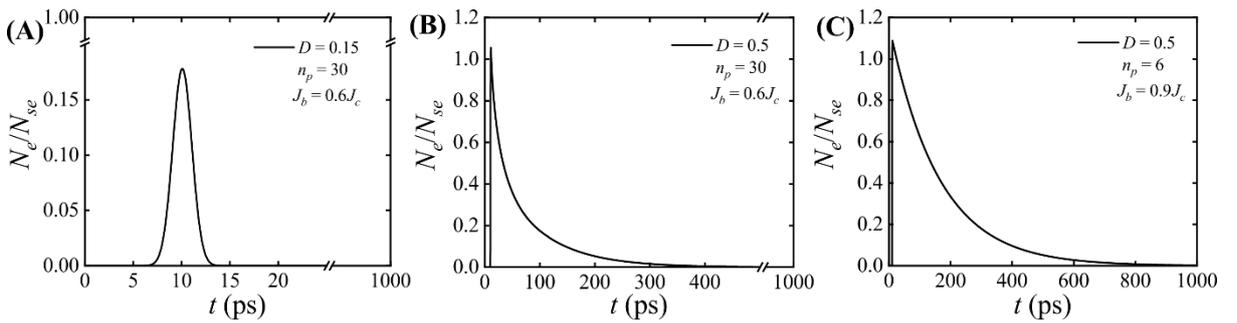

**Figure 05:** Variation of the ratio between excess energised electron and super-electrons, $N_e/N_{se}$, as a function of time for **(A)** $D = 0.15$, **(B)** $D = 0.5$ at $J_b = 0.6J_c$, $T_{sub} = 0.9T_{sw}$, and under irradiation of 1550 nm gaussian photon pulse ($n_p = 30$), and (C) $D = 0.5$ at $J_b = 0.9J_c$, $T_{sub} = 0.9T_{sw}$, and under irradiation of 1550 nm gaussian photon pulse ($n_p = 6$).

In **Fig. 05**, we have shown the variation of $N_e/N_{se}$ as a function of time, $t$ for two different level of dirtiness $D = 0.15$ (**Fig. 05(A)**) and 0.5 (**Fig. 05(B)**) under similar number photon ($n_p = 30$) irradiation. The cleaner SNSPD ($D = 0.15$) has almost 1.5 times higher $N_{se}$ than the cleaner SNSPD ($D = 0.5$) at $T_{sub} = 0.9T_{sw}$. The maximum value of $N_e/N_{se}$ is found to be 0.18 and 1.05 for $D = 0.15$ and $0.5$ respectively. Therefore, $n_p = 30$ is not high enough to form normal hotspot in NbN SNSPD for $D = 0.15$; whereas in $D = 0.5$ it breaks the superconductivity locally and we observe a detectable voltage pulse. Smaller $\varDelta$ and $D_e$ plays the major role in improving the sensitiveness of the dirtier SNSPD. We have also calculated $N_e/N_{se}$ for the dirty NbN SNSPD ($D = 0.5$) by biasing the device at $J_b/J_c = 0.9$ and $T_{sub} = 0.9T_{sw}$. The variation of $N_e/N_{se}$ as a function of time has been shown in **Fig. 05(C)**. Therefore, in conjunction with smaller $\varDelta$ and $D_e$, the higher Joule's heating observed due to higher normal state resistance promotes the improved sensitivity of the dirtier SNSPD.

The photon resolution and faster reset time are the two major performance characteristics of a SNSPD. While faster SNSPDs are essential in time-resolved measurements, high-speed quantum communication, better photon resolution is important for quantum, lidar and astronomical imaging where optical signals are very weak. Our model predicts that the SNSPD can be developed to operate into two different technological regimes by engineering the dirtiness and varying biasing conditions. We have thus proposed a generalized numerical model to quantify the photon sensitivity and reset time of practically dirty SNSPDs. We have shown with the increase in dirtiness, the NbN SNSPD can resolve a lower number of photons under the same biasing condition. We have calculated minimum number of photons ($n_p^*$) to create detectable hotspot and reset times ($\tau_{reset}$) for different biasing conditions ($J_b/J_c$) and several order of dirtiness ($D$). $n_p^*$ and $\tau_{reset}$ are found to be decreasing and increasing monotonically as dirtiness increases at constant $J_b$. Single photon sensitiveness has been observed in dirtiest NbN SNSPD having 115 times slower $\tau_{reset}$ suggesting trade-off between $n_p^*$ and $\tau_{reset}$.

## Acknowledgement

The authors acknowledge Defence Research and Development Organization (DRDO), Government of India, and Industrial Research and Development Unit, Indian Institute of Technology Delhi, India for financial support. The authors also acknowledge the members of Neos Lab, Department of Material Science and Engineering, Indian Institute of Technology Delhi, India for meaningful academic discussions.